# Superresolution observed from evanescent waves transmitted through nano-corrugated metallic films

Y. BEN-ARYEH

Physics Department Technion-Israel Institute of Technology, Haifa 32000, Israel
Fax:972-4-8295755, E-mail:phr65yb@physics.technion.ac.il

**ABSTRACT**  Plane EM waves transmitted through nano-corrugated metallic thin films produce evanescent waves which include the information on the nano-structures. The production of the evanescent waves at the metallic surface are analyzed. A microsphere located above the metallic surface collects the evanescent waves which are converted into propagating waves. The equations for the refraction at the boundary of the microsphere and the use of Snell's law for evanescent waves are developed. The magnification of the nano-structure images is explained by a geometric optics description, but the high resolution is related to the evanescent waves properties.

# 1    Introduction

Any microscopic image can be magnified by the use of a microscope. However, observing sub-wavelength structures with microscopes is difficult because of the *Abbe diffraction limit* [1], by which light with a wavelength $\lambda$ traveling in the medium with a refractive index $n$ and angle $\theta$ will make a spot with a radius

$$d = \frac{\lambda}{2(n\sin\theta)} . \qquad (1)$$

The term $n\sin\theta$ appearing in the denominator is called the numerical aperture (NA) and Abbe limit for ordinary microscopes is of order $\lambda/2$. To increase the resolution one may use UV and X-ray microscopes which increase the resolution due to their shorter wavelengths. Such microscopes suffer from lack of contrast in biological systems, are expensive and may also damage the sample.

The use of evanescent waves to increase the resolution beyond the Abbe limit can be related to Helmholtz eqution [2]. In homogeneous medium this equation is given as

$$(nk_0)^2 = k_x^2 + k_y^2 + k_z^2 , \qquad (2)$$

where $k_0 = (2\pi)/\lambda_0$, $\lambda_0$ is the wavelength in vacuum (or approximately in air), $n$ is the index of refraction (assumed it to be approximately real), and $k_x, k_y, k_z$ are the wavevector components. There are different mechanisms of producing evanescent waves by which we obtain

$$(k_x^2 + k_y^2) > (nk_0)^2 \qquad (3)$$

and then $k_z$ becomes imaginary ,i.e., there is a decay of the wave in the $z$ direction. The increase of the components of the wavevector $\vec{k}$ in the $(x,y)$ plane decreases the "effective" value of the wavelength in this plane, and thus increase the resolution. But the evanescent waves decay, however, in the $z$ direction perpendicular to the objects plane, so that in order to "capture" the fine structures which are available in the evanescent waves we need to put detectors very near the plane from which the electromagnetic (EM) waves are propagating.

Recently a very high resolution which is much better than the Abbe limit has been observed in microspheres, including the imaging by a microscope [3-5]. Although these results are quite impressive and has many applications, there is not yet a clear theoretical explanation for these phenomena. In this connection I would like to emphasize that although the magnification of the image by a microscope is explained by a geometric optics approach [3-5], such magnification will not increase the resolution unless there is an additional effect which increase the resolution beyond the Abbe limit. In previous studies [2,6-9] general mechanisms of increasing the resolution by using evanescent waves have been analyzed by using various methods. I would like to use these methods to analyze in the present paper the special properties of the microsphere system by which the evanescent waves lead to extremely high resolution.

One should take into account that EM evanescent waves incident with a large incidence angle are not transmitted through the microsphere due to total reflection [5]. Then, the microsphere acts as a thick lens producing the image of the corrugated metallic surface near the second focal plane, which is magnified afterward according to a conventional description of a microscope [10]. Our interest in the present paper is, however, in the analysis of the high resolution obtained by evanescent waves.

The present paper is arranged as follows:

In Section 2 the evanescent waves emitted from nano-curved metallic film by transmitted plane EM field is analyzed by using Fourier optics methods [2,6-9]. In Section 3 the transformation of evanescent waves into propagating waves by the microsphere is treated. The refraction of the evanescent waves at the boundary between the microsphere and the air is analyzed by following simple boundary equations for refraction at this boundary. The use of Snell's law for evanescent waves is developed. In Section 4 we summarize our results and conclusion, including a certain discussion.

## 2   Evanescent waves emitted from the surface of a thin metallic film, with nano-curved structures by transmitted plane EM field

The basic property of the microsphere systems is related to the development of laser nano-fabrication methods [11], by which nano-grooves on the metallic surfaces can be obtained. There are two mechanisms by which evanescent waves can be emitted from such surfaces: a) Plane EM waves incident on metallic thin film will produce transmitted wave in which the optical path through the metallic film will be variable according to the metallic nano-structures. There is a lot of literature on the optics of thin films (see e.g. [12]) but such analyses do not treat the optics of a thin film in which the optical path is a variable function of distances, in the transverse direction, in distances which are smaller than a wavelength. The new techniques of writing nano-scale patterns on thin metallic film enable the production of such thin films [11]. We expect that due to this effect the transmitted radiation will be a function of the transversal coordinates and such variations would lead to evanescent waves[2, 6,7,13]. b) Another mechanism for the production of evanescent waves from metallic surfaces can be related to plasmonic waves [14]. Surface plasmons are produced by TM EM waves which have a component of the electric field perpendicular to the metallic surface [14]. In the microsphere experiments the curved metallic surfaces have an electric field perpendicular to these surfaces (both for TM and TE waves). This effect would lead to *localized* surface plasmons [15] which increase the amount of evanescent waves.

Concerning the phenomena of obtaining high resolution with microspheres the combination of the above two effects might be important. It might be possible to check experimentally if the high resolution effects can be obtained also in corrugated non-metallic surfaces, e.g. by laser nano-fubrication in glass [16], for which there are not any plasmonic effects. It might be also possible to check experimentally if the high resolution effects are dependent on the kind of metal which is used as the surface plasmons intensity depends on the kind of the used metal. By following such experiments it might be possible to find what effect is the dominant one in the microsphere experiments. Although near field properties of microspheres have been treated [17], I would like to point out that *macroscopic* properties of plasmonic waves would not be enough to explain superresolution effects. Such effects should follow from *local* variations of the EM field as function of transversal distances which are smaller than a wavelength.

We can generalize the analysis following from the above effects (*localized surface plasmons* [15] and/or *localized* variation of optical path distances) by assuming that the EM waves in the air plane surface $z = 0$ which is a little beyond the metallic surface have a field distribution $E(x, y)_{z=0}$, where its fine structure is correlated with the metallic nano-structures.

Let us assume that the *scalar* EM field in the plane $z = 0$ can be represented as the following Fourier integral [2,6,7,13,18]:

$$E(x, y, z = 0) = \int_{-\infty}^{\infty} \int_{-\infty}^{\infty} \varepsilon(u,v) \exp[i(ux + vy)] du dv \quad , \tag{4}$$

where $u$ and $v$ are the *spatial coordinates* produced by the electric field in the $(x, y)$ plane, *including its fine structure*. The assumption of a scalar field [18-20] simplifies very much the analysis and the main properties of the evanescent waves can be related to such field.

For simplicity of analysis we assume that the microsphere is located symmetrically above the metallic film and is separated from its surface by a thin film of air. By using a straight forward analysis [2,6,7,13] we find that the transmitted electric field *in the air* before the microsphere is given by

$$E(x, y, z > 0) = \int_{-\infty}^{\infty} \int_{-\infty}^{\infty} \varepsilon(u,v) \exp[i(ux + vy + wz)] du dv \quad , \tag{5}$$

where

$$w^2 = k_0^2 - u^2 - v^2 \quad . \tag{6}$$

For cases for which $k_0^2 > u^2 + v^2$, $w_0$ becomes real and we get *propagating* waves in the $z$ direction. For cases for which $k_0^2 < u^2 + v^2$, $w$ becomes imaginary and we get evanescent waves decaying in the $z$ direction. The fine structure details of the metallic film are included in the evanescent waves.

For evanescent waves there is not any flow of energy in the propagation direction of the evanescent waves. This result [7] follows from the fact that for evanescent waves there is a phase difference of $\pi/2$ between the electric and magnetic fields, in the plane perpendicular to the propagation of the evanescent waves [1]. Therefore the Poynting vector in the propagation direction of the evanescent waves has zero time average. This conclusion is obtained also for a superposition of evanescent waves since usually there is not any phase correlation between the different components of the evanescent waves [2,6-9]. The role of the microsphere is to convert the evanescent waves into propagating waves.

The use of Eqs. (4-6) as the starting point for our analysis has the following advantages: a) It includes any possible mechanism for producing EM field in the $z = 0$ plane (a little beyond the metallic surface) which will be correlated with the metallic fine structure. b) Since the EM field emitted from the metallic film is given by a superposition of *plane waves* propagating in different directions the analysis for the refraction at the boundary of the microsphere becomes relatively simple

In Eq. (5) we have included only the EM field propagating into the air as we are assuming that the microsphere collects the transmitted waves, and we have not taken into account in this equation the *reflected* waves going back into the metal. A similar analysis to that made in the present article can be made for cases in which the microsphere collects the reflected EM wave from the metallic film [5].

## 3 Conversion of evanescent EM waves into propagating EM waves at the boundary between the air and the microsphere

We are interested in the region which is near to the contact point between the microsphere and the object plane, and in the conversion of evanescent waves into propagating waves. Then, the width $h$ of the thin film of air between the objects plain and the microsphere can be derived by using the relation

$$r^2 + (R-h)^2 = R^2 \quad , \tag{7}$$

where R is the radius of the microsphere, $r$ is the horizontal distance from the contact point, and $h$ is given at this distance. Eq. (7) leads to the relation

$$r^2 - 2Rh + h^2 = 0 \quad . \tag{8}$$

Since $2R \gg h$ we can neglect $h^2$ and get the approximate result

$$h = \frac{r^2}{2R} \quad . \tag{9}$$

We assume that $h \leq \lambda$ so that the evanescent waves have not decayed much before they are incidenting on the microsphere. For getting some orders of magnitude let us assume for example [3,4], $R = 4\mu m$, $\lambda = 5000 A^0$, then the circle for which the evanescent waves have not yet decayed, is given approximately by $\sqrt{2\lambda R} = 2\mu m$, which is not neglible relative to $R$.

According to Eq. (5) the scalar electric field at air on the boundary of the microsphere is given by

$$E(x,y,z=h) = \int_{-\infty}^{\infty} \int_{-\infty}^{\infty} \varepsilon(u,v) \exp\left[i(ux+vy+wh)\right] du dv \quad , \tag{10}$$

where

$$x^2 + y^2 + (R-h)^2 = R^2 \quad ; \quad x^2 + y^2 = r^2 \quad , \tag{11}$$

and $h$ is related approximately to $r$ by Eq. (9). Eq. (10) is described by integral over plane waves, where each plane wave is given as $\varepsilon(u,v)\exp[i(ux+vy+wh)]$ and the condition for this wave to be evanescent or propagating, respectively, in the air is given according to Eq. (6) by:

$$u^2 + v^2 > k_0^2 \Rightarrow evanescent\ wave\ ,$$
$$u^2 + v^2 < k_0^2 \Rightarrow propagating\ wave\ . \quad (12)$$

For the refracted EM plane wave in the microsphere, at the boundary, we get the relation

$$u'^2 + v'^2 + w'^2 = (k_0 n)^2 \quad , \quad (13)$$

where $w'$ is the component of the refracted wave at the boundary of the microsphere which is perpendicular to its surface, and $u'$ and $v'$ are the components parallel to its surface. $w'$ is imaginary (real) for evanescent (propagating) wave. According to Maxwell equations for a plane wave with a wave vector $\vec{k} = u\hat{x} + v\hat{y} + w\hat{z}$, in air, incident on the microsphere at a certain boundary point, the components of the wave vector which are parallel to surface of the microsphere are preserved. (Such relation is equivalent to the use of Snell' law which is valid also for an evanescent plane wave). I find two limiting cases in which evanescent waves are converted into propagating waves using corresponding approximations:

a) For plane waves incident very near to the contact point of the microsphere i.e. when $r \ll R$ we can use the approximation that the surface of the microsphere is parallel to the object plane. Then, the wave vector components $u'$ and $v'$ in the microsphere which are parallel to its surface satisfy the equalities

$$u' \cong u\ ;\ v' \cong v \quad , \quad (14)$$

and under the condition that the microsphere material has a real index of refraction $n$, the condition for evanescent and propagating wave in the microsphere is changed to

$$u'^2 + v'^2 \cong u^2 + v^2 > k_0^2 n^2 \Rightarrow evanescent\ wave\ ,$$
$$u'^2 + v'^2 \cong u^2 + v^2 < k_0^2 n^2 \Rightarrow propagating\ wave\ . \quad (15)$$

We find that evanescent waves in air which satisfy the relation $u^2 + v^2 > k_0^2$ become propagating wave in the microsphere if they satisfy the relation $u'^2 + v'^2 < k_0^2 n^2$, so that many evanescent waves are converted to propagating waves.

b) For a plane wave incident on the microsphere for which the condition $r \ll R$ is not valid (although $r < R$) the analysis becomes even more favorable to the conversion of evanescent waves into propagating waves. Due to the spherical symmetry of the microsphere we can reduce the problem to refraction of a plane EM wave propagating in the $(x, z)$ plane in air, with a wave vector $\vec{k}_0 = u\hat{x} + w\hat{z}$. Such representation can be made *for each plane wave* as we can choose the $x$ coordinate so that it will be in the direction of the sum of wave vector components in the object plane. Due to this description the analysis of refraction and reflection at the boundary of the microsphere is reduced to corresponding equations in the $(x, z)$ plane. Under the condition $u^2 > k_0^2$ such waves are evanescent in the air. We assume that the plane wave is incident on the microsphere at a point $O$ located by a vector $\vec{R}$, relative to the center of the microspere, and that $\theta$ is the angle between this vector and the vertical one.

We are interested in the conversion of EM evanescent waves to propagating waves into the microsphere, since the information on the fine structures of the corrugated metallic surface is included in the evanescent waves in the air. According to Eq. (10) an evanescent wave $\varepsilon(\mu)$, $(\mu^2 > k_0^2)$ which is incident at point $O$ of the microsphere is given by the EM field

$$E(x, z = h) = \varepsilon(u) \exp\left[i(ux + wh)\right] \quad , \quad (16)$$

where $w$ is imaginary, and we have reduced the analysis to a plane $(x, z)$ by using the spherical symmetry of the microsphere. Then, the components of the vector $\vec{k} = u\hat{x} + w\hat{z}$ which are parallel and perpendicular to the surface of the microsphere at point $O$ at air, denoted by tilde, are given by

$$\begin{aligned} \vec{\tilde{u}} &= u\cos\theta \hat{x} + w\sin\theta \hat{z} \quad , \quad parallel \\ \vec{\tilde{w}} &= w\cos\theta \hat{x} - u\sin\theta \hat{z} \quad , \quad perpendicular \end{aligned} \quad , \quad (17)$$

where $\hat{x}$ and $\hat{z}$ are the unit vectors in the $x$ and $z$ directions, respectively. We notice that this transformation satisfy the relation

$$\vec{\tilde{u}}^2 + \vec{\tilde{w}}^2 = u^2 + w^2 = k_0^2 \quad . \quad (18)$$

We should take care of the fact that we treat here evanescent wave for which $w$ is imaginary. Due to the continuity equation the component of the wavevector in the microsphere, denoted by prime, which is parallel to its surface at the point $O$ is given by

$$\vec{u}' = \vec{\tilde{u}} = u\cos\theta\,\hat{x} + w\sin\theta\,\hat{y} \qquad . \tag{19}$$

The components of the wave vector $\vec{k}'$ in the microsphere satisfy the relation

$$u'^2 + w'^2 = (k_0 n)^2 \tag{20}$$

We find that the evanescent waves in air are transformed into propagating waves if they satisfy the relation

$$u^2 \cos^2\theta - |w|^2 \sin^2\theta \leq k_0^2 n^2 \quad (propagating\ waves\ in\ microsphere) \quad . \tag{21}$$

These waves remain evanescent in the microsphere if they satisfy the relation

$$u^2 \cos^2\theta - |w|^2 \sin^2\theta > k_0^2 n^2 \quad (evanescent\ waves\ in\ microsphere) \ . \tag{22}$$

Since for a large values of $w$ and a relatively large value for $\theta$

$$u^2 \cos^2\theta - |w|^2 \sin^2\theta \ll u^2 \qquad , \tag{23}$$

we find that relation (21) can improve the condition for the transformation of evanescent waves into propagating waves by an order of magnitude relative to the condition $u^2 < k_0^2 n^2$ which is valid under the condition $\theta \approx 0$.

We should take into account that Eqs. (16-22) have been obtained in the $(x, z)$ plane assuming that the sum of the components of the wave vector in the object plane is in the $x$ direction. Similar equations will be obtained for any $(\tilde{x}, z)$ plane (in case the sum of the wave vector components in the object plane is in the $\tilde{x}$ direction) where the $(\tilde{x}, z)$ plane is obtained from the $(x, z)$ plane by a rotation around the $z$ axis. Such property for the boundary equations are related to the spherical symmetry of the microsphere. *The values of $u$ and $w$* are, however, different for different planes as they depend on the image of the metallic corrugated object, which is usually non-symmetric. Using the spherical symmetry of the microsphere the condition (21) for propagating waves can be generalized to the three dimensional case as

$$(u^2 + v^2)\cos^2\theta - |w|^2 \sin^2\theta \leq k_0^2 n^2 \quad (propagating\ waves\ in\ microsphere), \tag{24}$$

where $u$ and $v$ are the components of the wave vector in the x and y coordinates, respectively, and $w$ is in the $z$ direction assuming it to be imaginary.

We have analyzed here the plane EM waves properties by which evanescent waves are converted into propagating waves. The intensity of each transmitted wave is decreased by the reflectance properties of the microsphere which can be analyzed by conventional methods, but are of less interest here. For large values of $\theta$ the resolution is increased but the width $h$ of the air film also increases which leads to a decrease in the transmittance intensity of evanescent waves due to a decay which is of order $\exp(-|w|h)$.

For using Snell's law for evanescent waves in air we find according to Eqs. (17-20):

$$\sin\theta = \frac{|\tilde{u}|}{k_0} = \frac{\sqrt{u^2\cos^2\theta - |w|^2\sin^2\theta}}{k_0}, \qquad (25)$$

$$\sin\theta' = \frac{|\tilde{u}|}{nk_0} = \frac{\sqrt{u^2\cos^2\theta - |w|^2\sin^2\theta}}{nk_0}. \qquad (26)$$

Eqs. (25-26) have been developed for the plane $(x,z)$ assuming that the total component of the wavevector in the object plane is given by $u$ in the $x$ direction. In the general case that we have two components $u$ and $v$ of the wavevector in the $x$ and $y$ directions, respectively, Eqs. (25-26) are generalized as

$$\sin\theta = \frac{\sqrt{(u^2+v^2)\cos^2\theta - |w|^2\sin^2\theta}}{k_0} \qquad (27)$$

$$\sin\theta' = \frac{\sqrt{(u^2+v^2)\cos^2\theta - |w|^2\sin^2\theta}}{nk_0} \qquad (28)$$

We find that Snell's law $\sin\theta = n\sin\theta'$ is valid. However, when Eq. (26) or Eq. (28) would lead to the result $\sin\theta' > 1$ the evanescent wave will remain evanescent also in the microsphere, which is consistent with the above analysis.

## 5 Summary, discussion and conclusion

In previous studies [2,6-9] it has been shown that the convolution between the spatial modes of the evanescent waves and the spatial modes of the *tip detector* leads in the *near field analysis* to conversion of evanescent waves into propagating waves. In this connection I would like to explain that the microsphere used for getting high resolution acts as a tip-detector *which leads to tunneling of the evanescent waves* into the microsphere, where there they are converted into propagating waves. It has been shown in the present research that there are two effects which can lead to the tunneling of evanescent waves: a) By using boundary conditions we find that since the microsphere has an index of refraction $n > 1$ which is larger than that of air ($n \simeq 1$), a part of the evanescent waves are converted into propagating waves. This effect is the dominant one only for tunneling near the contact point between the microsphere and the metallic surface ($\sin\theta \ll 1$). b) Using the spherical geometry of the microspheres which are very small (diameters of order of some microns) and assuming a certain geometry to the metallic thin film corresponding to its nano-structures, it has been shown that the boundary conditions lead to a very strong tunneling of evanescent waves into propagating waves for relatively large values of $\theta$. This effect is analogous to that discussed in previous works by a convolution description but the knowledge of the microsphere geometry allowed us to get more explicit results by using Fourier optics and boundary conditions *for the spatial modes*.

There are two sources for the evanescent waves: 1) Local variations of optical path lengths in the transversal direction which are much smaller than a wavelength. 2) Local surface plasmons which change the field intensities in the corrugated metallic surfaces including the information on the nano-structures. It has been possible to combine these two effects by assuming spatial modes in the plane above the metallic surface which correspond to the nano-structures. A general analysis for the evanescent waves produced by plane waves transmitted through corrugated metallic thin film due to these two effects has been given in Section 2.

In Section 3 the refraction of the evanescent waves at the boundary of the microsphere has been analyzed. Using the spherical geometry of the microsphere, boundary equations for the EM field components which are parallel and perpendicular to the surface of the microsphere have been developed. The conditions by which evanescent waves are converted into propagating waves in the refraction of the EM field are described. The use of Snell's law for evanescent waves has been developed.

In conclusion, although the optical geometry description for the magnification of nano-metallic structure images by microshere and microscope is valid [3-5], for interpreting the high resolution obtained, it is necessary to make a different analysis for the conversion of evanescent waves, into propagating waves.